\documentclass[aps,pre,reprint,superscriptaddress]{revtex4-2}
\usepackage{booktabs,graphicx,siunitx}
\usepackage[italicdiff]{physics}
\DeclareMathOperator*{\argmax}{\arg\!\max}
\DeclareMathOperator{\Var}{Var}
\usepackage{xcolor,hyperref}
\hypersetup{
colorlinks,
linkcolor={blue!50!black},
citecolor={blue!50!black},
urlcolor={blue!80!black}
}
\begin{document}
\title{Johari-Goldstein $\beta$ relaxation in glassy dynamics originates from two-scale energy landscape}
\author{Kumpei Shiraishi}
\email{kumpeishiraishi@g.ecc.u-tokyo.ac.jp}
\author{Hideyuki Mizuno}
\affiliation{Graduate School of Arts and Sciences, University of Tokyo, Komaba, Tokyo 153-8902, Japan}
\author{Atsushi Ikeda}
\affiliation{Graduate School of Arts and Sciences, University of Tokyo, Komaba, Tokyo 153-8902, Japan}
\affiliation{Research Center for Complex Systems Biology, Universal Biology Institute, University of Tokyo, Komaba, Tokyo 153-8902, Japan}
\date{\today}
\begin{abstract}
Supercooled liquids undergo complicated structural relaxation processes, which have been a long-standing problem in both experimental and theoretical aspects of condensed matter physics.
In particular, past experiments universally observed for many types of molecular liquids that relaxation dynamics separated into two distinct processes at low temperatures.
One of the possible interpretations is that this separation originates from the two-scale hierarchical topography of the potential energy landscape; however, it has never been verified.
Molecular dynamics simulations are a promising approach to tackle this issue, but we must overcome laborious difficulties.
First, we must handle a model of molecular liquids that is computationally demanding compared to simple spherical models, which have been intensively studied but show only a slower process: $\alpha$ relaxation.
Second, we must reach a sufficiently low-temperature regime where the two processes become well separated.
Here, we handle an asymmetric dimer system that exhibits a faster process: Johari-Goldstein $\beta$ relaxation.
Then, we employ the parallel tempering method to access the low-temperature regime.
These laborious efforts enable us to investigate the potential energy landscape in detail and unveil the first direct evidence of the topographic hierarchy that induces the $\beta$ relaxation.
We also successfully characterize the microscopic motions of particles during each relaxation process.
Finally, we study the predictive power of low-frequency modes for two relaxation processes.
Our results establish for the first time a fundamental and comprehensive understanding of experimentally observed relaxation dynamics in supercooled liquids.
\end{abstract}
\maketitle

\textbf{When liquids are cooled while avoiding crystallization, their dynamics dramatically slow down.
This is a famous, major problem in the glass transition phenomena.
Vast experiments universally observed two processes of slow dynamics: a faster Johari-Goldstein (JG) $\beta$ relaxation and a slower $\alpha$ relaxation.
However, most past theoretical and numerical works have intensively focused on the $\alpha$ relaxation, whereas only limited attention has been given to the JG $\beta$ relaxation.
Here, we perform extensive MD simulations on an asymmetric dimer system, which shows both relaxation processes.
We report the first detailed analyses of JG $\beta$ relaxation, including the real-space dynamics of molecules and hierarchical potential energy landscape.
Our results contribute to closing a gap that has existed for a long time between theoretical and experimental sides of the research on glass transition phenomena.}

\vspace{1.5em}

Complex relaxation dynamics emerge when liquids are cooled toward the glass transition temperature $T_g$~\cite{Ediger1996,Angell2000,Ngai_2011}.
Understanding the origins and characterizing the properties of relaxation processes are of fundamental importance in condensed matter physics.
Theoretical and numerical studies on glassy dynamics~\cite{Berthier_Biroli_2011} usually treat the system with simplifications of realistic features, e.g., shapes of molecules.
In return, and unfortunately, a phenomenon has been totally overlooked: the Johari-Goldstein (JG) $\beta$ relaxation.

At low temperatures near $T_g$, the faster JG $\beta$ relaxation branches off from a primary slower $\alpha$ relaxation process~\cite{Ediger1996,Angell2000,Ngai_2011}.
Originally, this fast process was observed in polymers and attributed to the internal degrees of freedom of the constituents.
However, Johari and Goldstein discovered that this process appeared even in rigid molecules with no intramolecular degrees of freedom~\cite{Johari_Goldstein_1970}, and JG $\beta$ relaxation is now considered intrinsic to the glass transition phenomena as a precursor to $\alpha$ relaxation~\cite{Ngai_1998}.
Furthermore, since $\alpha$ relaxation freezes first, JG $\beta$ relaxation is the main relaxation process at temperatures below $T_g$.

Various aspects of this faster process have been investigated, including firm classification criteria of the genuine JG $\beta$ process~\cite{Ngai_JCP_2004}, shapes of the relaxation spectrum (excess wing or $\beta$ peak)~\cite{Schneider_2000,Ngai_2001}, temperature dependence of the relaxation time~\cite{Paluch_2003}, and decoupling from $\alpha$ relaxation~\cite{Saito_2012}.
Additionally, experimental studies recently note that JG $\beta$ relaxation is closely related to theoretical notions such as the Gardner transition and mosaic state~\cite{Geirhos_2018,Caporaletti2021}.
It is now considered that JG $\beta$ relaxation is a universal property that is widely shared by molecular glasses regardless of the type of constituent~\cite{Thayyil_2008}.
In addition to molecular glasses, metallic glasses also show JG $\beta$ relaxation~\cite{Yu_review_2013,Yu_2014,Wang_MG_2019}.

There are fewer theoretical and numerical works on JG $\beta$ relaxation, which are markedly in contrast with many more experimental works.
Of those that have been done, the most important idea is based on the potential energy landscape (PEL)~\cite{Goldstein_1969,Debenedetti_2001}.
In his pioneering paper, Stillinger noted that the two-scale hierarchical structure in the PEL is reasonable to interpret the two distinct relaxation processes~\cite{Stillinger_1995}.
According to his explanation, the inter-basin transitions correspond to JG $\beta$ relaxation, whereas the transitions between structures consisting of multiple basins, which is called \textit{metabasin}, correspond to the $\alpha$ relaxation.
Although his study of metabasin was at the conceptual level, numerical studies of the PEL were vigorously pursued by Heuer and his colleagues~\cite{Heuer_2008}.
Heuer also suggested that the concept of metabasin was useful to understand the mechanism of JG $\beta$ relaxation~\cite{Vogel_2004}.
Additionally, mean-field replica theory recently argued that JG $\beta$ relaxation could be interpreted as \textit{sub-basin} transitions in a metabasin~\cite{Charbonneau_fractal_2014}.

However, no study has successfully demonstrated the relationship between JG $\beta$ relaxation and PEL.
Molecular dynamics (MD) simulations have been an indispensable tool to study the PEL since the earliest attempt by Stillinger and Weber~\cite{Stillinger_1982}.
It is desirable to detect JG $\beta$ relaxation in MD simulations not only from PEL perspective but also from real-space perspective because MD simulations give complete information of the microscopic motions of molecules and can provide insight into the origin of the process.
However, to do this, we must overcome two laborious difficulties as follows.
(i) It is computationally demanding to equilibrate the system at very low temperatures where JG $\beta$ relaxation is separated from the $\alpha$ relaxation, and most importantly,
(ii) JG $\beta$ relaxation does not clearly appear even when spherical particles, which have been intensively studied by the vast majority of previous works, are cooled at very low temperatures.

In 2012, Fragiadakis and Roland successfully detected JG $\beta$ relaxation in MD simulations as a distinct peak in the relaxation spectrum using a model of asymmetric dimer particles~\cite{Fragiadakis_2012}.
However, their analysis focused on the heterogeneous dynamics~\cite{Fragiadakis_2014}, and analyses on the microscopic dynamics during the JG $\beta$ relaxation and the hierarchy of the PEL remain absent.
In addition, their low-temperature data are limited to the aging dynamics.

\begin{figure}
\centering
\includegraphics[width=\linewidth]{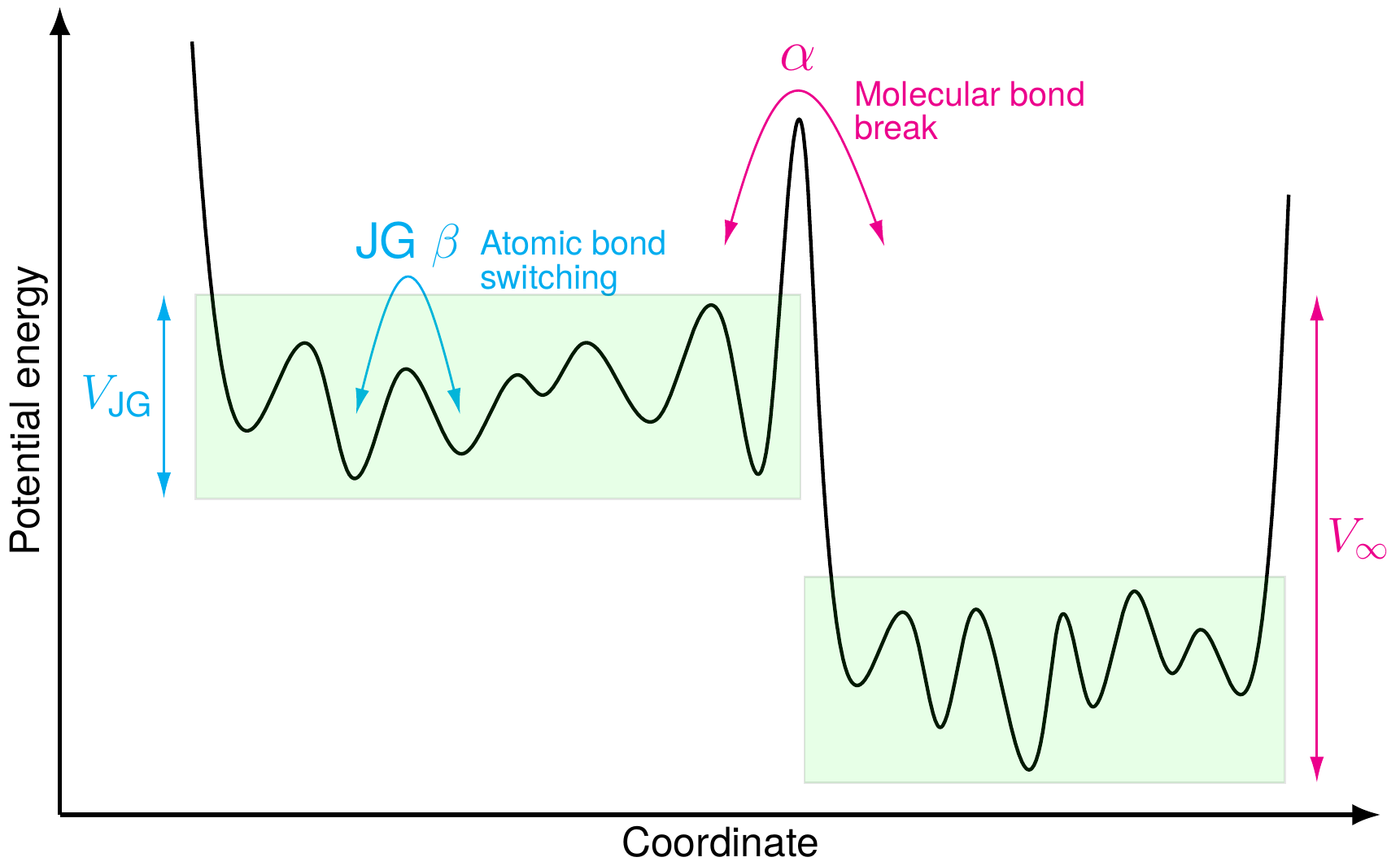}
\caption{Schematic summary of this paper.
A newly introduced quantity, which is a variance of time series of the inherent structure potential energy, shows the hierarchical structure of topography.
From a real-space perspective, the switching of atomic bonds corresponds to JG $\beta$ relaxation.
Correlation between low-frequency vibrational modes and the subsequent relaxation persists for the $\alpha$ relaxation time.}
\label{fig:cartoon}
\end{figure}

Here, we perform extensive MD simulations on asymmetric dimers and report detailed analyses on JG $\beta$ relaxation from the perspectives of both real-space dynamics and PEL.
First, by applying the parallel tempering (PT) method~\cite{Hukushima_1996,Yamamoto_2000} to this system for the first time, we access the equilibrium dynamics at very low temperatures, which had never been previously accessed.
Then, we introduce the analysis of bond-break correlation in the dimer system and elucidate the microscopic processes for both JG $\beta$ and $\alpha$ relaxations.
Furthermore, we report the first direct evidence of two intrinsic structures of PEL in the system that clearly show the $\beta$ peak in the relaxation spectrum.
Finally, we confirm that the correlation between low-frequency vibrational modes and relaxations holds as in spherical particle systems and persists for a timescale of $\alpha$ relaxation, although the topography of the energy landscape is different in the dimer system.
A schematic summary of our analysis is shown in Fig.~\ref{fig:cartoon}.

\section*{Three-step relaxation}
\begin{figure*}[th!]
\centering
\includegraphics[width=\linewidth]{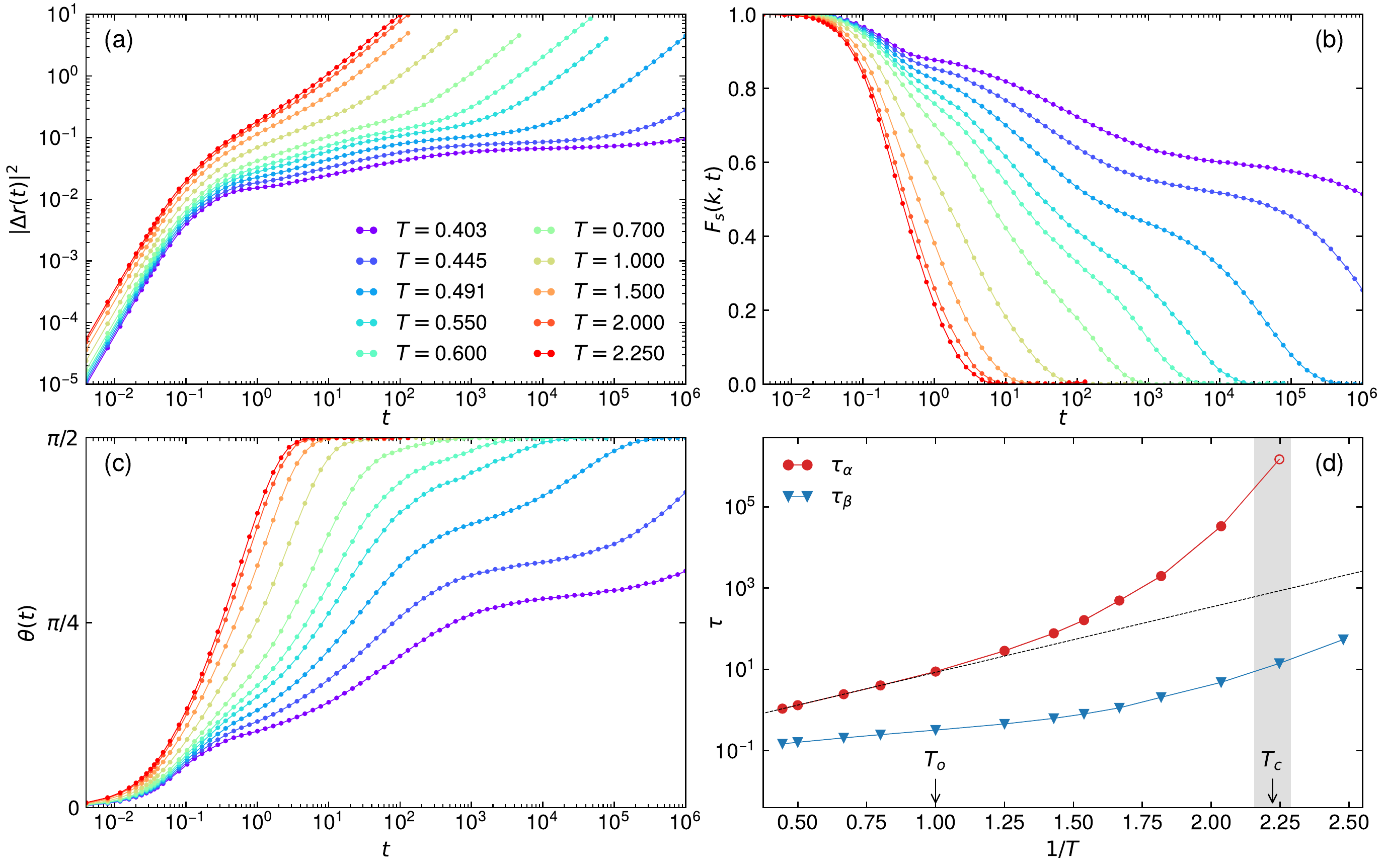}
\caption{(a) Mean-squared displacement $\abs{\Delta\vb*{r}(t)}^2$ and (b) self-part of the intermediate scattering function $F_s(k, t)$ of the center of mass of dimers.
(c) Angle $\theta(t)$ between the direction vectors at times $t$ and $0$ of the dimers.
(d) Arrhenius plot of relaxation times for $\alpha$ and JG $\beta$ relaxation.
For $T = 0.445$, we obtained $\tau_\alpha$ by extrapolating the linear fit of the final four points of $F_s(k,t)$ to the threshold.
Therefore, we express the point by an open symbol.
Arrhenius behavior is fitted by the four highest-temperature points of $\tau_\alpha$ and depicted in the figure by the dashed line.
The onset temperature is $T_o = 1.0$, where $\tau_\alpha$ deviates from Arrhenius behavior.
The mode-coupling temperature $T_c \approx 0.45$ is detected in Fig.~S1.
All data in this figure are the quantities for the AB dimers.}
\label{fig:dynamics}
\end{figure*}

The translational dynamics of dimers are investigated by calculating the mean-squared displacement (MSD) of the center of mass of each dimer $\abs{\Delta\vb*{r}(t)}^2$; see Fig.~\ref{fig:dynamics} (a).
At all temperatures, MSD in the short timescale shows a power-law behavior with an exponent of 2 ($\abs{\Delta\vb*{r}(t)}^2 \propto t^2$), i.e., ballistic motion.
At high temperatures, this ballistic motion smoothly crosses over to the diffusive motion of $\abs{\Delta\vb*{r}(t)}^2 \propto t$, where a plateau regime is observed between these two motions.
This behavior is similar to that of spherical models such as the Kob-Andersen (KA) model~\cite{Kob_Andersen_I_1995}.
The striking difference emerges at low temperatures, where we observe a three-step behavior with two plateaus.
For example, at $T = 0.491$, MSD shows one plateau at $t \sim 10^0$ at a height of $2 \times 10^{-2}$ and the other plateau at $t \sim 10^3$ at a height of $10^{-1}$.

Equivalently, the self-part of the intermediate scattering function $F_s(k,t)$ at low temperatures shows the salient feature of the three-step relaxation; see Fig.~\ref{fig:dynamics} (b).
For example, at $T = 0.491$, $F_s(k,t)$ starts to decay at $t \sim 10^{-1}$ due to vibrations and exhibits a shoulder at $t \sim 10^0$.
Then, the decay due to a faster JG $\beta$ relaxation expands a relatively larger time window until $t \sim 10^2$, and another plateau appears.
The final decay of a slower $\alpha$ relaxation starts at $t \sim 10^4$.

We define the characteristic times of $\alpha$ and JG $\beta$ relaxations by $F_s(k, \tau_\alpha) = 0.2$ and $F_s(k, \tau_\beta) = 0.75$, respectively, and show their temperature dependence in Fig.~\ref{fig:dynamics} (d).
$\tau_\alpha$ clearly shows a super-Arrhenius temperature dependence, whereas $\tau_\beta$ shows a milder increase with decreasing temperature.
The dashed line in Fig.~\ref{fig:dynamics} (d) exhibits the Arrhenius behavior fitted from the four highest-temperature points, and $\tau_\alpha$ deviates from the Arrhenius behavior (line) at $T_o = 1.0$.
We find that the potential energy of inherent structures starts to decay from a plateau at this onset temperature $T_o = 1.0$, which is identical to those in the sphere models without the JG $\beta$ process~\cite{Sastry_1998}.
The mode-coupling temperature of the present system is $T_c \approx 0.45$, which is estimated as a temperature where the fitted line of $\tau_\alpha^{-1/\gamma}$ for several $\gamma$ crosses zero; see Fig.~S1.
Using the power of the parallel tempering protocol, we can present the relaxation time below the mode-coupling crossover, which hampers the sampling of the conventional MD.

We also calculate the relaxation spectrum by performing the Fourier transform with respect to time $t$ on the data of $F_s(k,t)$ (Fig.~S2) and confirm that at low temperatures, it shows two distinct peaks in the frequency regime, which correspond to $\tau_\alpha$ and $\tau_\beta$ and merge at high temperatures.
Thus, our numerical simulations of the dimer system well reproduce the experimental observations of relaxation dynamics.

The rotational dynamics are observed by the time evolution of $\theta(t)$, which is the angle between the direction vectors at times 0 and $t$.
See \textit{Materials and Methods} for the definition.
In Fig.~\ref{fig:dynamics} (c), $\theta(t)$ also shows a three-step behavior with two plateaus, similar to MSD.
However, the height of the second plateau after JG $\beta$ relaxation (at $t \sim 10^3$) is quite large (more than $\pi/4$) compared to the height of the plateau in MSD at the same timescale, which is approximately $10^{-1}$.
These results indicate that JG $\beta$ relaxation corresponds to a rotational motion with large changes in the directions of dimers but few changes in their translational positions.

\section*{Microscopic characterization of relaxations}
\begin{figure}
\centering
\includegraphics[width=\linewidth]{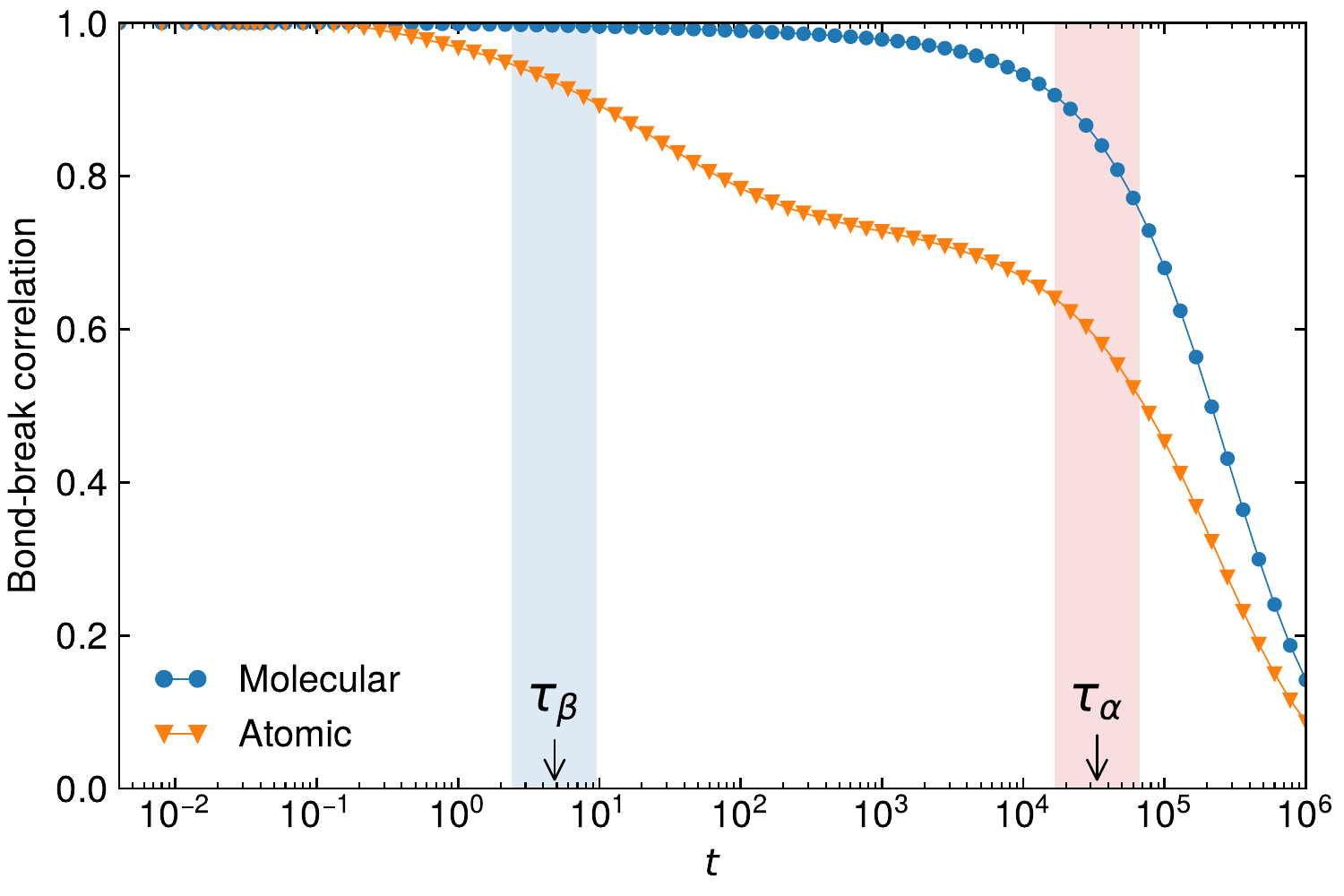}
\caption{Molecular and atomic bond-break correlation at $T = 0.491$. The characteristic times from Fig.~\ref{fig:dynamics} (d) are also shown.}
\label{fig:BB}
\end{figure}

To more microscopically characterize the relaxation processes, we introduce the bond-break correlation~\cite{Yamamoto_1998}.
This quantity has been well studied in spherical particles, is known to be insensitive to the motion in the plateau regime of MSD, and starts to decay at a longer timescale than $\tau_\alpha$ defined by $F_s(k,t)$~\cite{Shiba_2012}.
Note that a recent study shows that the relaxation times defined from the bond-break correlation and $F_s(k,t)$ become identical at temperatures much below $T_c$~\cite{Scalliet2022}.
First, we calculate the molecular bond-break correlation at $T = 0.491$ (Fig.~\ref{fig:BB}).
See \textit{Materials and Methods} for the definitions and parameters in this analysis.
Evidently, the correlation does not show diminishment at the timescale of JG $\beta$ relaxation and starts to decay when $F_s(k,t)$ reaches $\tau_\alpha$.
This behavior is qualitatively identical to that of spherical particles.
This result signifies that the $\alpha$ relaxation of dimers corresponds to the rearrangement of the molecular bond network.

Next, we introduce the atomic bond-break correlation to examine whether the atomic bonds that compose a molecular bond are changed during JG $\beta$ relaxation even where the molecular bond-break correlation is invariant.
The definition and parameters are given in \textit{Materials and Methods}.
Figure~\ref{fig:BB} shows that the atomic bond-break correlation decreases during JG $\beta$ relaxation.
Thus, unlike molecular bonds, the atomic bond network changes, and atoms lose their environmental information during the JG $\beta$ relaxation process.

Let us scrutinize this change in atomic bonds by classifying the bonds between dimer molecules.
Since our model of dimers has an asymmetry, i.e., a dimer molecule is composed of a large atom and a small atom, we can categorize atomic bonds into three types: between two large atoms, between a large atom and a small atom, and between two small atoms.
Furthermore, considering the number and types of atomic bonds that compose a molecular bond, we obtain a total of 12 states of molecular bonds that completely classify all molecular bonds; see \textit{Materials and Methods} for details and Table~\ref{bond_states} for a list of states of molecular bonds.

Following the above classification, we analyze the states of all molecular bonds taken during their time evolution and find that remarkably, more than \SI{99}{\percent} of the molecular pairs dissociate through a specific path of states.
The graphical illustration of this path is shown in Fig.~\ref{fig:bond_image}.
For a detailed description of the state transitions, see Supplementary Information.
We note that most pairs return to the state that they left once, but once dissociated, it is rare to form the bond again.

As described in Fig.~\ref{fig:bond_image}, JG $\beta$ relaxation corresponds to the switching of atomic bonds, and it has a rotational motion that does not greatly change the translational position.
This aspect of JG $\beta$ relaxation is consistent with the findings in the previous section, where we assess the translational and rotational dynamics by MSD and $\theta(t)$.
We also note that this microscopic aspect of JG $\beta$ relaxation as local bond switching has also been reported in experimental studies of a metallic glass~\cite{Liu_2014} and a hydrogen-bond glass~\cite{Caporaletti_2019}.

Meanwhile, the $\alpha$ relaxation corresponds to the breaking of the molecular bond because the molecular bond-break correlation decays after the relaxation (Fig.~\ref{fig:BB}).
Therefore, the $\alpha$ relaxation can be interpreted as the cut of the last remaining bond between the large atoms (Fig.~\ref{fig:bond_image}).
By combining the insight from MSD, this cut of the atomic bond and the concomitant molecular bond break are attributed to cage-breaking events that drastically change the local environments.

\begin{figure}
\centering
\includegraphics[width=\linewidth]{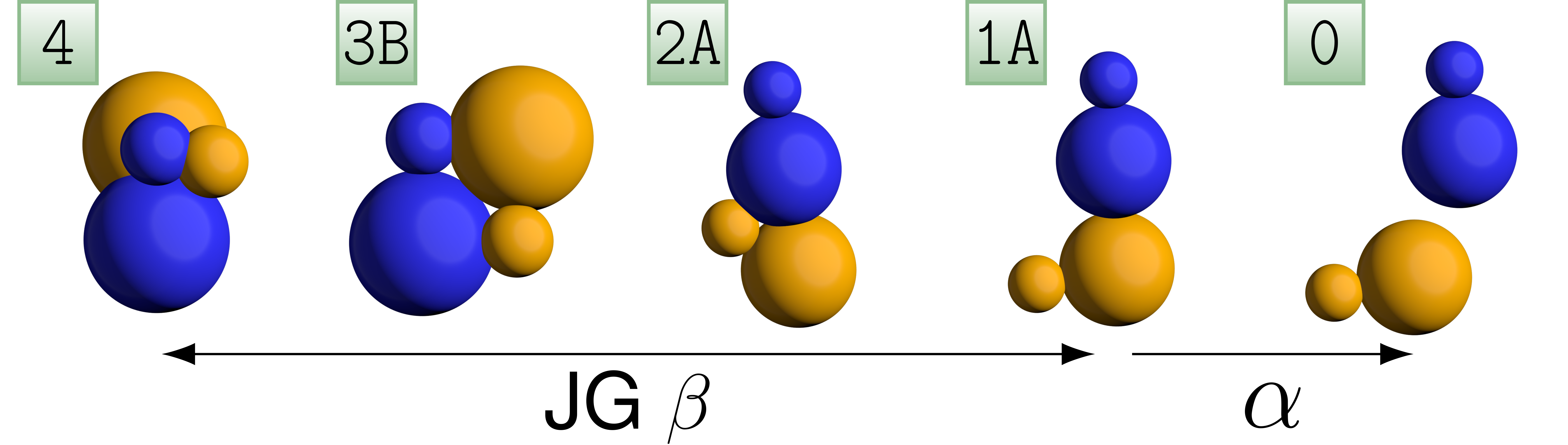}
\caption{Graphical illustration of the main dissociation process.
Each label in a box corresponds to a classification in Table~\ref{bond_states}.
The molecular bond goes back and forth from state \texttt{4} to state \texttt{1A} during JG $\beta$ relaxation.
The relative positions of the two dimers are altered by rotational motions, which cause the switching of atomic bonds.
The last remainder, which is the LL bond, is cut in the $\alpha$ relaxation.}
\label{fig:bond_image}
\end{figure}

\section*{Two-scale energy landscape}
\begin{figure*}[th!]
\centering
\includegraphics[width=\linewidth]{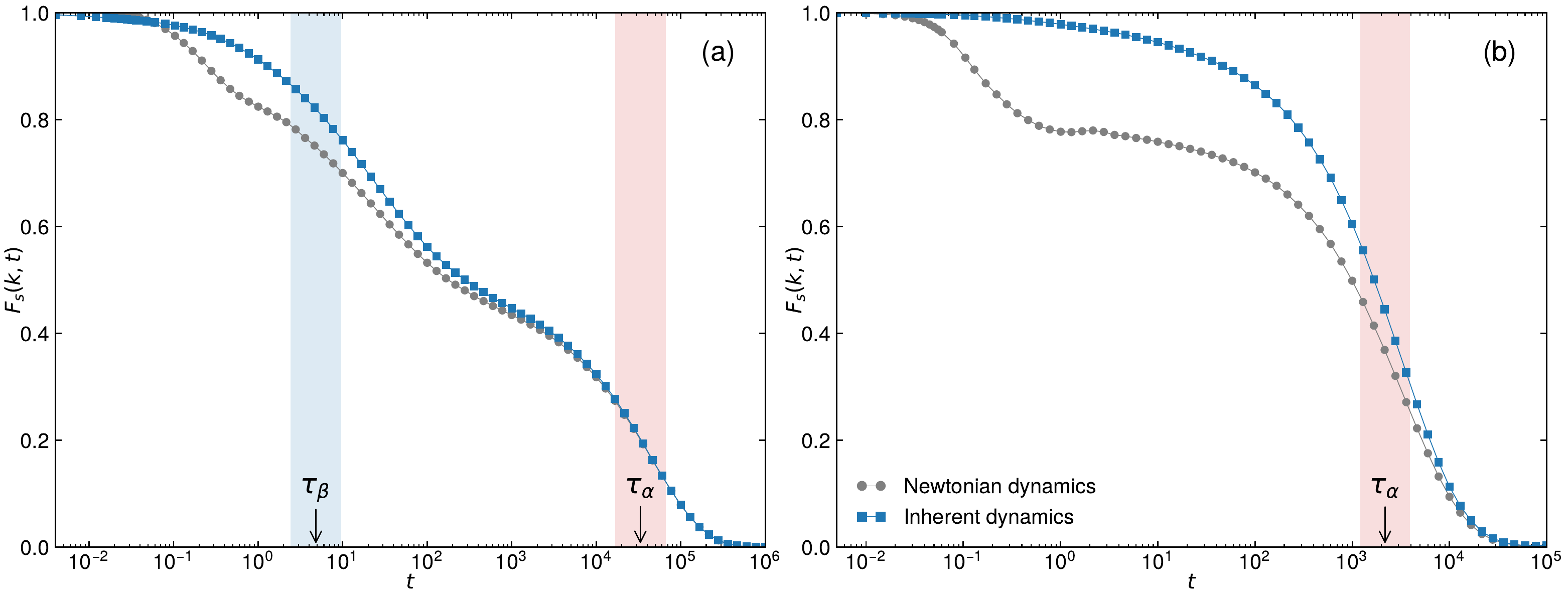}
\caption{Comparison between $F_s(k,t)$ in inherent dynamics and Newtonian dynamics for (a) the dimer system of $T=0.491$ and (b) KA system of $T=0.45$.}
\label{fig:ISdynamics}
\end{figure*}

We now turn our attention to the PEL.
To study the PEL, we calculate the inherent structures (IS), which are configurations of particles with the system potential being locally minimized (configurations in the zero-temperature limit).
First, we consider a time series of inherent structures that correspond to a trajectory of Newtonian dynamics.
We call them IS trajectories in the inherent dynamics.
We calculate $F_s(k,t)$ for both dimer system and KA system in both inherent and Newtonian dynamics (see Fig.~\ref{fig:ISdynamics}; see Supplementary Information for the simulation details of the KA system).
First, for the KA system, $F_s(k,t)$ of the inherent dynamics shows a one-step decay, as observed in the previous study~\cite{Schroder_2000}.
The significant decay starts at approximately the same timescale as that of $\tau_\alpha$.
The first decay of $F_s(k,t)$ in Newtonian dynamics corresponds to intra-basin vibrations, so it does not affect the inherent dynamics.
Therefore, the dynamics of the KA system suggest that it stays inside a basin before the $\alpha$ relaxation.

Meanwhile, for the dimer system, $F_s(k,t)$ of the inherent dynamics behaves much differently from that of the KA system.
The relaxation starts on a much shorter timescale than $\tau_\alpha$, which is roughly the same timescale as $\tau_\beta$.
This behavior is in stark contrast to that of the KA system.
Afterwards, a plateau appears before $\tau_\alpha$, which is in close agreement with $F_s(k,t)$ of Newtonian dynamics.
This short-time deviation from $F_s(k,t) = 1$ of the inherent dynamics indicates that there are considerable inter-IS transitions on the timescale of $\tau_\beta$.
In addition, the appearance of a plateau in $F_s(k,t)$ of the inherent dynamics can be interpreted as an exploration for a PEL structure that is not sufficient with only inter-IS transitions.
This result supports Stillinger's two-scale PEL interpretation for JG $\beta$ relaxation.

\begin{figure}
\centering
\includegraphics[width=\linewidth]{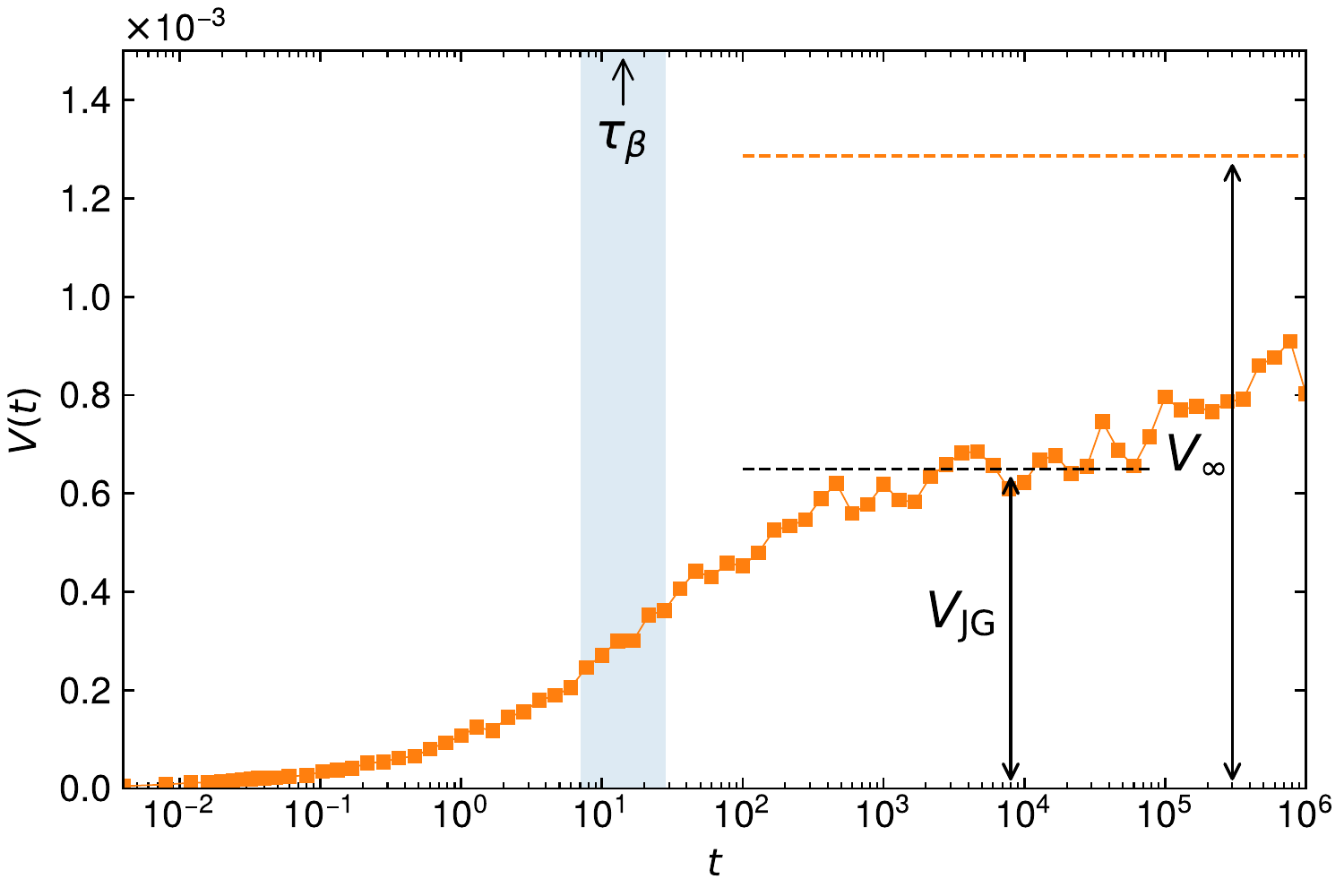}
\caption{Variance $V(t)$ of potential energies at $T = 0.445$.
$V(t)$ reaches a plateau $V_\text{JG}$ after JG $\beta$ relaxation.
It increases again toward a long-time limit value $V_\infty$.
$V_\infty$ cannot be observed for the current simulation time.
See the discussion and data of other temperatures in the Supplementary Information.}
\label{fig:variance}
\end{figure}

Here, to more explicitly demonstrate the hierarchical structure, we introduce a new analysis of the PEL.
We introduce a quantity $V(t)$:
\begin{align}
 V(t) = \Var[E_\text{IS}(t) - E_\text{IS}(0)],
\end{align}
where $E_\text{IS}(t)$ is the potential energy of a configuration in an IS trajectory at time $t$.
$V(t)$ is the time-dependent variance of the IS energies, so it quantifies the range of the PEL explored until time $t$.
In Fig.~\ref{fig:variance}, we show $V(t)$ at $T = 0.445$ and see that $V(t)$ increases over time and reaches a plateau after $\tau_\beta$.
We express the height of the plateau as $V_\text{JG}$ (dashed line).
The time $t \sim 10^4$ that $V(t)$ reaches $V_\text{JG}$ coincides with the time that $F_s(k,t)$ reaches the second plateau (Fig.~\ref{fig:dynamics} (b)).
At this timescale, the range of the PEL explored by the trajectory is limited by $V_\text{JG}$.
This is reminiscent of the concept of metabasin, where the system is confined before the $\alpha$ relaxation.

As time evolves, $V(t)$ increases again, which implies that the system explores a more global topography of the potential energy landscape.
In the long-time limit, $V(t)$ converges to a finite value $V_\infty$, which should be the variance of the inherent structure energy at given temperatures.
At $T = 0.445$, this is not reached due to the limited length of our simulation time (Fig.~\ref{fig:variance}).
However, as shown in Fig.~S4, we confirm that $V(t)$ does converge to $V_\infty$ at a high temperature ($T = 1.500, 0.800$).
We estimate $V_\infty$ from $V(t)$ at $t \gtrsim \tau_\alpha$, shown as the horizontal dashed lines in Fig.~S4.
Additionally, at temperatures lower than $T = 0.445$, the system cannot explore the PEL beyond $V_\text{JG}$ in the current simulation time, and $V(t)$ persists on the plateau at a height of $V_\text{JG}$ (see Supplementary Information).
Thus, we expect that exploration of the metabasin can be completed on longer timescales than $\tau_\alpha$.
This is consistent with recent works of spherical particles that demonstrate the existence of broad distributions behind the metabasin transitions~\cite{Baity-Jesi_2021}, dynamics of supercooled liquids~\cite{Berthier_2021}, and excess-wing spectrum of relaxations~\cite{Guiselin_2022}.

In summary, our calculation of $V(t)$ clarifies that the system traverses subbasins within a metabasin near the initial IS during JG $\beta$ relaxation.
This metabasin is composed of many ISs whose energy variance is $V_\text{JG}$.
Then, at longer times, the system reaches the global metabasin exploration during the $\alpha$ relaxation (see also Fig.~\ref{fig:cartoon}).
Thus, to the best of our knowledge, we provide the first numerical demonstration of Stillinger's picture of two-scale PEL~\cite{Stillinger_1995}.

\section*{Structural indicator for the relaxation}
Since the discovery by Widmer-Cooper and coworkers~\cite{Widmer_Cooper_2008}, it has been well known that the low-frequency vibrational modes of the initial IS (i.e.\ the IS quenched from the initial instantaneous configuration) are correlated with subsequent thermal relaxation.
Additionally, searching for a suitable ``structural indicator'' to predict the dynamic heterogeneity~\cite{Ediger_2000} is an active research topic~\cite{Schoenholz_2016,Tong_2018,Bapst_2020,Boattini_2020,Paret_2020}.
As we have seen, frequent inter-basin transitions occur during JG $\beta$ relaxation, and metabasin transitions occur during the $\alpha$ relaxation in the dimer system.
Here, we investigate whether the low-frequency modes can still predict relaxation dynamics in this two-scale landscape and, if so, to what timescales it persists.

To do this, we first perform vibrational mode analysis to obtain data on the low-frequency modes of this system (see \textit{Materials and Methods} for the formulation of the vibrational mode analysis).
It is now well established that amorphous systems with spherical potential have so-called quasi-localized vibrational modes in the low-frequency regime~\cite{Lerner_2016,Mizuno_2017}.
In our previous works~\cite{Shiraishi_2019,Shiraishi_2020}, we also found that quasi-localized modes emerged in dimer systems.
Interestingly, these quasi-localized modes show rotational motions of particles.
We will report more detailed information regarding the vibrational properties of dimer systems in our next paper.

Next, we consider the correlation between low-frequency modes and relaxations.
As predictors, we use the norm of eigenvectors on each molecule/atom at the initial IS.
As actual relaxations, we adopt the propensity of motion, $\theta_i(t)$ of each molecule, and the atomic bond-break correlation of each atom.
Then, we quantify the correlation between these relaxations at time $t$ and the vibrations using three evaluators: the Rank~\cite{Patinet_2016}, Pearson, and Spearman correlations.
See \textit{Materials and Methods} for details of the calculation.
For the correlations with the atomic bond-break correlation, the eigenvectors of each molecule were converted to the eigenvectors of each atom~\cite{Shiraishi_2020}, and their norms were considered predictors.

\begin{figure}
\centering
\includegraphics[width=\linewidth]{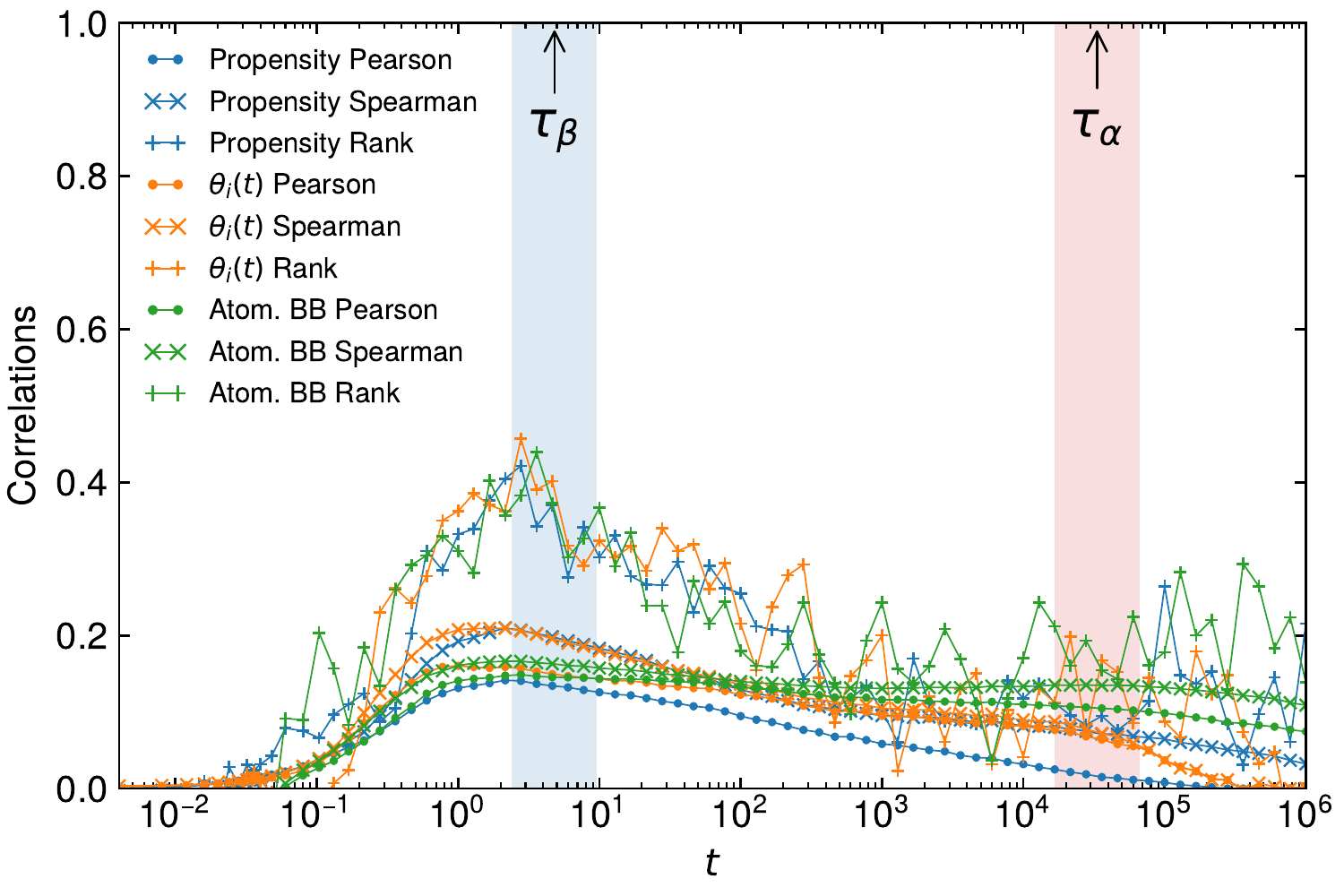}
\caption{Correlations between vibrations and relaxations for dimers at $T = 0.491$.
Relaxations are measured in the propensity of motion, $\theta_i(t)$, and atomic bond-break correlation.}
\label{fig:predict}
\end{figure}

In Fig.~\ref{fig:predict}, we show the correlations of all possible cases at $T = 0.491$.
We see that the correlations persist for the $\alpha$ relaxation.
The value of the correlation seems to be lower than that of the KA system (see Fig.~S5).
We attribute this result to the introduction of subbasins associated with JG $\beta$ relaxation, which act as obstacles and deteriorate correlations.
The prediction power nonetheless holds for long-time relaxations.
This observation is surprising since multiple basin transitions occur due to JG $\beta$ relaxation before traversing metabasins due to the $\alpha$ relaxation.
Therefore, although the system experiences transitions between many subbasins during JG $\beta$ relaxation, the low-frequency modes quantified at the initial state (local minimum) can predict the $\alpha$ relaxation that succeed the $\beta$ relaxation (see Fig.~\ref{fig:cartoon}).

\section*{Discussion and outlook}
In conclusion, we have performed extensive MD simulations on the dimer system, an archetypical model that exhibits JG $\beta$ relaxation in addition to $\alpha$ relaxation, and provided a comprehensive understanding of the relaxation processes of supercooled molecular liquids.
We first confirm that the scattering function shows a three-step decaying behavior with a fast JG $\beta$ relaxation and a slow $\alpha$ relaxation; thus, our simulation results reproduce well experimental observations on molecular liquids.
Next, we introduce the concepts of molecular bonds and atomic bonds and study the relaxation processes based on the microscopic motions of particles.
Remarkably, simulations demonstrate that more than \SI{99}{\percent} of molecular pairs follow a specific path of bond breakage.
In particular, we unveil that the JG $\beta$ relaxation is the declination process of atomic bonds due to the rotational motions of molecules; then, the $\alpha$ relaxation is that of the cut of the last remaining bond, which implies the molecular bond breakage and cage-breaking event.

Our main finding is that the dimer system explores a two-scale landscape of potential energy.
Our analysis of the time-dependent variance of the IS energies shows that during the JG $\beta$ relaxation, the system explores a more detailed landscape, i.e., subbasins within a large metabasin, whereas the $\alpha$ relaxation is the exploration process of a more global landscape, i.e., metabasins.
Thus, we establish that molecular liquids have a hierarchical landscape with two characteristic scales, whereas atomic liquids show only one energy scale.
Remarkably, the low-frequency modes associated with a subbasin can predict the activation dynamics over the metabasins ($\alpha$ relaxation).
This result indicates that each of many subbasins within a metabasin retains global metabasin information.

Our results close a gap between the theoretical idea of the PEL and the experimental observations on relaxation dynamics.
There are several possible ways to develop in this direction.
First, fragile/strong properties of liquids are considered stages in which the PEL plays a major role~\cite{Debenedetti_2001}.
Strong liquids are thought to exhibit a single metabasin topography, while fragile liquids are thought to have multiple metabasins, but direct numerical evidence remains scarce.
This paper will provide another method to determine the difference in topography other than the disconnectivity graph~\cite{Middleton_2001}.
Second, it is also interesting to apply statistical-physics-based methods~\cite{Berthier_2014} to sample basins in a metabasin on a more global scale to identify the nature of basins that yields JG $\beta$ relaxation.
In the present study, we sample the basins that are visited in the path in Newtonian dynamics.
However, the hierarchical structure of the free energy landscape can be detected by calculating the Franz-Parisi potential, which is the free energy of glasses~\cite{Franz_Parisi_PRL_1997}.
This type of analysis of JG $\beta$ relaxation is still lacking.
In addition, inspired by theoretical progress in the mean-field limit~\cite{simple_glasses}, the Gardner transition at low temperatures has attracted much attention in recent years and been recently experimentally explored~\cite{Seguin_2016,Geirhos_2018,Hammond_2020,Albert_2021}.
It is an interesting question how the appearance of JG $\beta$ relaxation and the corresponding introduction of basins in the PEL interact with the splitting of basins in the deeply supercooled phase caused by this transition.
However, it is difficult to achieve equilibration at temperatures far below $T_c$ using the PT method, as in the present study.
Thus, it will be useful to develop a suitable system for efficient algorithms such as the swap Monte Carlo as it was done in spherical particles~\cite{Ninarello_2017}.

\section*{Materials and Methods}
\subsection*{Model}
We study a binary mixture of three-dimensional dimer molecules~\cite{Fragiadakis_2013}.
Each dimer comprises two atoms, and the distance $d$ between the two is fixed to 0.5.
There are AB and CD molecules in our system, where A, B, C, and D represent atomic types.
All atoms have an equal mass $m$.
Atoms of different molecules interact with the Lennard-Jones potential
\begin{align}
 \phi\pqty{r_{ij}} = 4\epsilon_{ij}\bqty{\pqty{\frac{\sigma_{ij}}{r_{ij}}}^{12} - \pqty{\frac{\sigma_{ij}}{r_{ij}}}^6}.
\end{align}
The interaction parameters are $\epsilon_\text{AA} = \epsilon_\text{AB} = \epsilon_\text{BB} = \epsilon$, $\epsilon_\text{CC} = \epsilon_\text{CD} = \epsilon_\text{DD} = 0.5\epsilon$ and $\epsilon_\text{AC} = \epsilon_\text{AD} = \epsilon_\text{BC} = \epsilon_\text{BD} = 1.5\epsilon$.
$\sigma_{ij}$ for pairs of identical types are $\sigma_\text{AA} = \sigma$, $\sigma_\text{CC} = 0.88\sigma$, $\sigma_\text{BB} = r\sigma_\text{AA}$, and $\sigma_\text{DD} = r\sigma_\text{CC}$, where $r = 0.5$ is a parameter for the asymmetry of dimers.
$\sigma_{ij}$ for pairs of different types are $\sigma_{ij} = S_{ij}\pqty{\sigma_{ii} + \sigma_{jj}}$; $S_{ij} = 0.5$ if $(i,j) = (\text{A}, \text{B}), (\text{C}, \text{D})$ and $S_{ij} = 0.4255$ if $(i,j) = (\text{A}, \text{C}), (\text{A}, \text{D}), (\text{B}, \text{C}), (\text{B}, \text{D})$.
Since the discontinuity of the pair force at the cutoff distance $r_\text{c} = 2.5\sigma_{ij}$ strongly affects the properties of the low-frequency modes~\cite{Shimada_LJ_2018}, we employ $V\pqty{r_{ij}} = \phi\pqty{r_{ij}} - \phi\pqty{r_\text{c}} - \pqty{r_{ij} - r_\text{c}}\phi^\prime\pqty{r_\text{c}}$ as an interaction potential to ensure the continuity of the force and the potential.
We consider $N_\text{mol} = 1000$ dimer molecules enclosed in a square box with periodic boundary conditions with $N_\text{AB} = 4N_\text{mol}/5$ and $N_\text{CD} = N_\text{mol}/5$.
The number density is $\rho = N_\text{mol}/L^3 = 1.25$.
Lengths, energies, temperatures, and time are measured in units of $\sigma$, $\epsilon$, $\epsilon/k_B$, and $\pqty{m\sigma/\epsilon}^{1/2}$, respectively.

\subsection*{Simulations}
We perform MD simulations in the \textit{NVT} ensemble using the Nos\'{e}-Hoover thermostat~\cite{Frenkel_Smit} with a time step of 0.004 to equilibrate the system.
To fix the distance between atoms during the simulation, we employ the RATTLE method combined with the velocity Verlet algorithm~\cite{Allen_Tildesley}.
For lower temperatures, we use the parallel tempering (PT) method~\cite{Hukushima_1996,Yamamoto_2000} to equilibrate the system.
In our PT simulation, 48 replicas are used, and each replica corresponds to a temperature ranging from 1.500 to 0.385.
Exchange trials are performed every 1250 MD steps using the Metropolis criterion.
We start the sampling of the configurations every 10000 exchange trials after 110000 trials.
We check the equilibrium by observing the lack of aging behavior in $F_s(k,t)$ of production runs.
Configurations from PT simulation are used only for $T = 0.491, 0.445, 0.403$.
After equilibration is performed, we start MD simulations in the \textit{NVE} ensemble for production runs.
The number of initial configurations is 95 for $T = 0.491, 0.445, 0.403$ and at least 100 for other temperatures.
To analyze the correlation between low-frequency modes and relaxations, we use the data in the isoconfigurational ensemble~\cite{Widmer_Cooper_2004} (the ensemble size is 24).
The FIRE algorithm~\cite{Bitzek_2006} is used for energy minimization.
The convergence of the algorithm is judged by whether the maximum value of the norms of the forces acting on each atom is less than $10^{-10}$.
All simulations are performed in an in-house code that uses MPI to handle the parallel computation.

\subsection*{Dynamic observables}
The mean-squared displacement is calculated as
\begin{align}
 \abs{\Delta\vb*{r}(t)}^2 = \ev{\frac{1}{N_\text{AB}} \sum^{N_\text{AB}}_{i=1} \abs{\vb*{r}_i(t) - \vb*{r}_i(0)}^2},
\end{align}
where $\vb*{r}_i(t)$ is the translational position of the center of mass of dimer $i$ at time $t$, and $\ev{}$ is the ensemble average.
The self-part of the intermediate scattering function is
\begin{align}
 F_s(k,t) = \ev{\frac{1}{N_\text{AB}} \sum^{N_\text{AB}}_{i=1} \cos \bqty{\vb*{k} \vdot \Delta\vb*{r}_i(t)}},
\end{align}
where $k=7.25$.
The rotational dynamics are studied by
\begin{align}
 \theta(t) = \ev{\frac{1}{N_\text{AB}} \sum^{N_\text{AB}}_{i=1} \acos \bqty{\vu*{d}_i(t) \vdot \vu*{d}_i(0)}},
\end{align}
where $\vu*{d}_i(t)$ is the unit direction vector of dimer $i$ at time $t$.

\subsection*{Bond state analysis}
To analyze the microscopic processes of JG $\beta$ relaxation, we calculate the bond-break correlation, which is the fraction of remaining neighbors since time 0~\cite{Yamamoto_1998}.
For the dimer system, we introduce two types of this quantity: molecular and atomic bond-break correlations.
In the calculation of molecular bond-break correlation, we consider the index of neighboring molecules.
Molecules are classified as connected if even one of the constituent atoms is closer than a threshold.
The molecular bond-break correlation $B^i_\text{m}(t)$ is formulated as
\begin{align}
 B^i_\text{m}(t) = \frac{n^i_\text{m}(t|0)}{n^i_\text{m}},
\end{align}
where $n^i_\text{m}$ is the number of molecules connected to molecule $i$ at time 0, and $n^i_\text{m}(t|0)$ is the number of remaining neighboring molecules of molecule $i$ at time $t$.
In the calculation of atomic bond-break correlation $B^i_\text{a}(t)$, we consider the index of neighboring atoms to calculate similarly.
We use the threshold values of $1.4\sigma_{ij}$ at time 0 and $1.7\sigma_{ij}$ at time $t$.

From the definition of molecular bonds, we can categorize the bonds into several cases.
First, the number of atomic bonds that compose a molecular bond can take the values 4, 3, 2, 1, or 0.
Furthermore, since molecules are composed of large atoms (A or C; written as L) and small atoms (B or D; written as S), a single atomic bond can take on the states of LL, LS, or SS, which lead to several branches.
As a result, molecular bonds can be categorized into 12 states, as summarized in Table~\ref{bond_states}.

\begin{table}
 \centering
 \caption{Molecular bond states}
 \begin{tabular}{cl}
  \toprule
  States      & Atomic bond breakdown \\
  \midrule
  \texttt{4}  & LL, LS, LS, SS \\
  \texttt{3A} & LL, LS, SS     \\
  \texttt{3B} & LL, LS, LS     \\
  \texttt{3C} & LS, LS, SS     \\
  \texttt{2A} & LL, LS	       \\
  \texttt{2B} & LS, SS	       \\
  \texttt{2C} & LS, LS	       \\
  \texttt{2D} & LL, SS	       \\
  \texttt{1A} & LL             \\
  \texttt{1B} & LS             \\
  \texttt{1C} & SS             \\
  \texttt{0}  & -              \\
  \bottomrule
 \end{tabular}
 \label{bond_states}
\end{table}

\subsection*{Correlation between vibrations and relaxations}
First, we perform the standard vibrational analysis at the initial IS to obtain eigenvalues and eigenvectors~\cite{Ashcroft_Mermin}.
By appropriately linearizing the equations of motion~\cite{Shiraishi_2020}, we formulate the dynamical matrix
\begin{align}
 \mathcal{M} = \pdv[2]{V}{\vb*{r}}{\vb*{r}},
\end{align}
where $\vb*{r} = \pqty{\vb*{r}_1, \dots, \vb*{r}_{N_\text{mol}}}$ is the generalized coordinates of this system (size $5N_\text{mol}$), and $V = \sum_{i,j}V(r_{ij})$ is the potential.
A detailed formulation and the explicit elements of the dynamical matrix can be found in our previous paper~\cite{Shiraishi_2020}.
Note that the mass is $2m$, and the moment of inertia is $d^2/2$ in the present analysis.
The eigenvalue problem of $\mathcal{M}$ is solved numerically using the SciPy package~\cite{2020SciPy-NMeth}.

Norms of eigenvectors on each dimer are used as the predictors for relaxations.
We sum the norm over the lower \SI{1}{\percent} of all eigenmodes (i.e.\ 50 lowest-frequency modes).
We quantify subsequent actual relaxations by the propensity of motion $\abs{\vb*{r}_i(t) - \vb*{r}_i(0)}$, $\theta_i\pqty{t} = \acos \bqty{\vu*{d}_i(t) \vdot \vu*{d}_i(0)}$ of each molecule, and the atomic bond-break correlation of each atom $1 - B^i_\text{a}(t)$.
Note that atomic bond-break correlations are subtracted from 1 to ensure that relaxed particles take a larger value.

To evaluate the correlation between vibrations and relaxations, we employ the Rank, Pearson, and Spearman correlations.
Here, we denote $p_i$ as the predicted relaxation and $t_i$ as the true relaxation of particle $i$.
In the calculation of the Rank correlation~\cite{Richard_plasticity_2020}, we first consider $i_\text{max} = \argmax_i t_i$ to evaluate $r_x$, the rank of $p_{i_\text{max}}$ normalized by the number of particles $N_\text{mol}$ ($2N_\text{mol}$ for the case of atomic bond break as $t_i$).
$r_x$ is sampled over initial configurations (95 samples) and a cumulative histogram with 25 bins is constructed with the dataset of $r_x$.
If a predictor has excellent prediction ability, the histogram will show a stepwise increase at $r_x = 0$.
If a predictor and a true relaxation are uncorrelated, the histogram will linearly increase.
Therefore, the Rank correlation is defined as $2A-1$, where $A$ is the area of the cumulative histogram, which ranges from 0 (poor) to 1 (excellent).
The Pearson correlation $\rho_p$ is defined as
\begin{align}
 \rho_p = \frac{\sum_i \pqty{p_i - \bar{p}} \pqty{t_i - \bar{t}}}{\sqrt{\sum_i (p_i - \bar{p})^2} \sqrt{\sum_i (t_i - \bar{t})^2}}.
\end{align}
The Spearman correlation is the Pearson correlation between the rank variables.
The NumPy~\cite{Harris2020} and SciPy~\cite{2020SciPy-NMeth} packages are used to calculate the Pearson and Spearman correlations, respectively.
We sweep $t_i$ for the time series and examine the time dependence of each correlation.

\section*{Author Contributions}
K.S., H.M., and A.I. designed research, performed research, analyzed data, and wrote the paper.

\begin{acknowledgments}
We thank Daniele Coslovich and Koji Hukushima for discussions.
This work is supported by JSPS KAKENHI (Grants Nos.~18H05225, 19H01812, 20H00128, 20H01868, 21J10021, 22K03543), and the Initiative on Promotion of Supercomputing for Young or Women Researchers, Information Technology Center, the University of Tokyo.
\end{acknowledgments}

\section*{Supplementary Information}
\subsection*{Mode-coupling temperature}
\begin{figure}
\centering
\includegraphics[width=\linewidth]{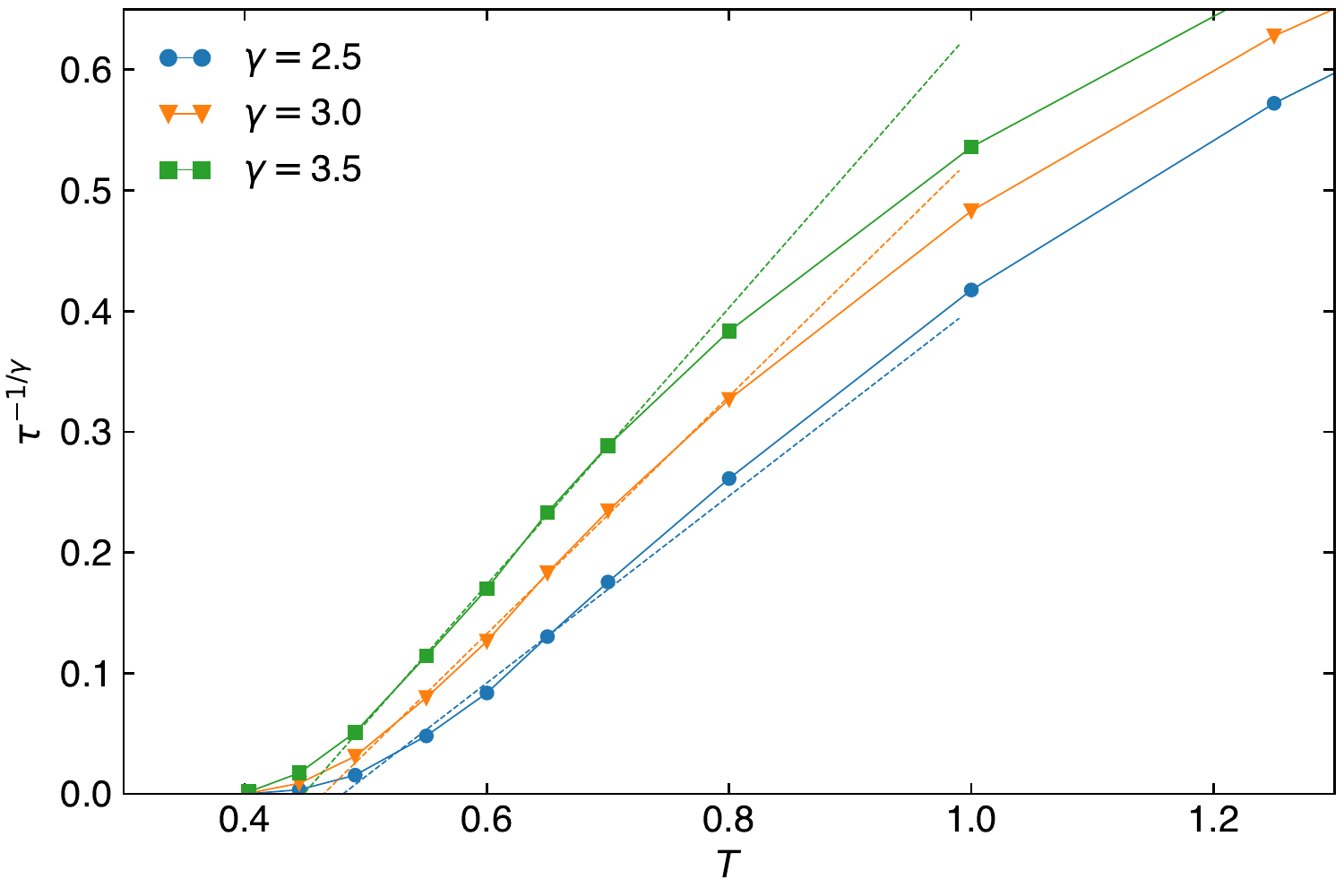}
\caption{$\tau_\alpha^{-1/\gamma}$ versus $T$ for $\gamma = 2.5, 3.0, 3.5$. For all cases, the fitted lines (the dashed lines) cross zero at around $T_c \approx 0.45$.}
\label{fig:Tc}
\end{figure}

We plot $\tau_\alpha^{-1/\gamma}$ versus $T$ for several $\gamma$ and observe that the mode-coupling temperature of the dimer system is $T_c \approx 0.45$ (Fig.~\ref{fig:Tc}).

\subsection*{Relaxation spectrum}
\begin{figure}
\centering
\includegraphics[width=\linewidth]{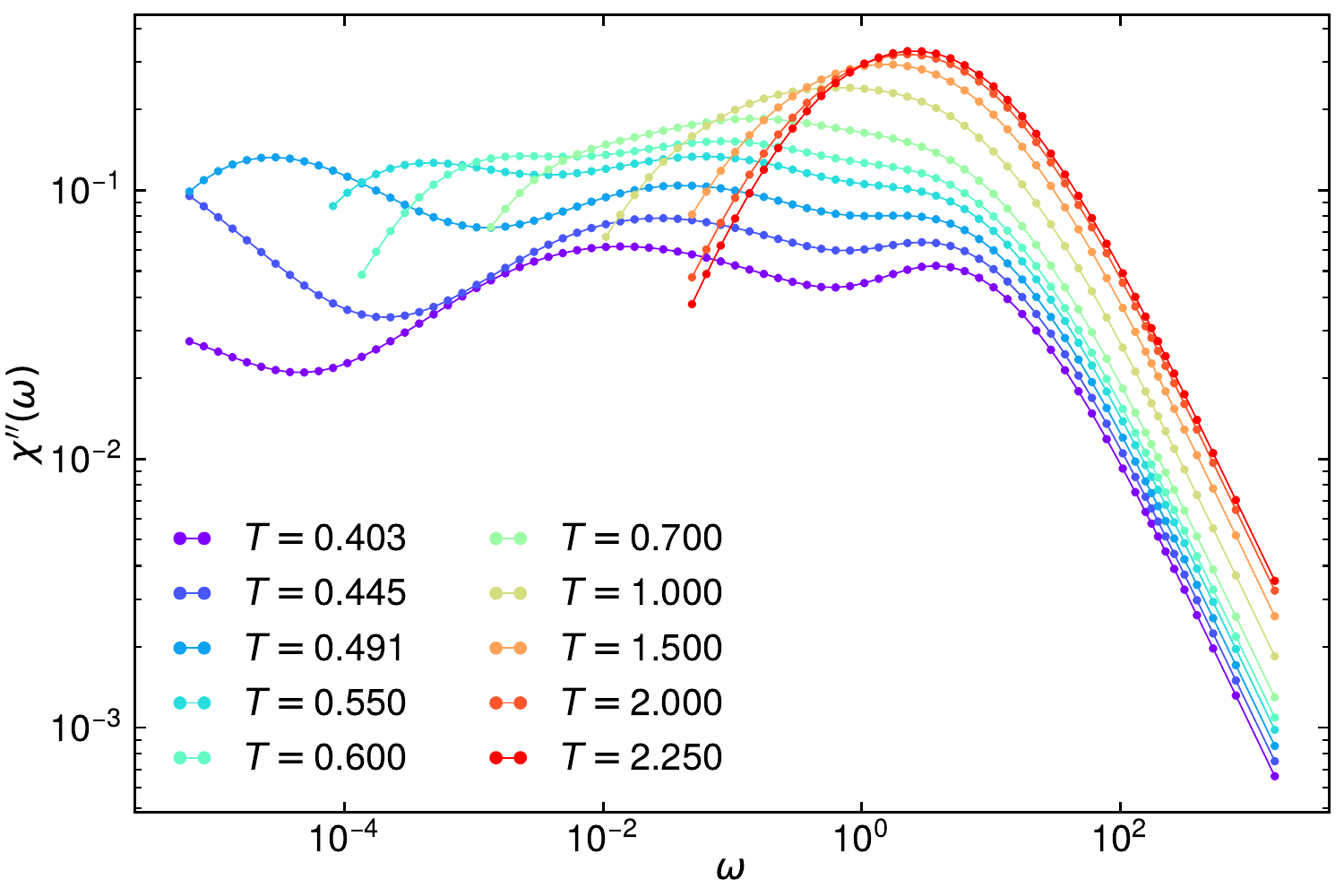}
\caption{Relaxation spectrum for each temperature.}
\label{fig:spectrum}
\end{figure}

The relaxation spectra $\chi^{\prime\prime}(\omega)$ is calculated as~\cite{Guiselin_2022}
\begin{align}
 \chi^{\prime\prime}(\omega) = -\sum_{k} \dv{F_s(t_k)}{\log t} \frac{\omega t_k}{1 + (\omega t_k)^2} \log \pqty{\frac{t_k}{t_{k-1}}},
\end{align}
where
\begin{align}
 \dv{F_s(t_k)}{\log t} = \frac{F_s(t_k) - F_s(t_{k-1})}{\log (t_k) - \log (t_{k-1})}.
\end{align}
The resulting spectrum is shown in Fig.~\ref{fig:spectrum}.

\subsection*{Bond state analysis}
\begin{figure}
\centering
\includegraphics[width=\linewidth]{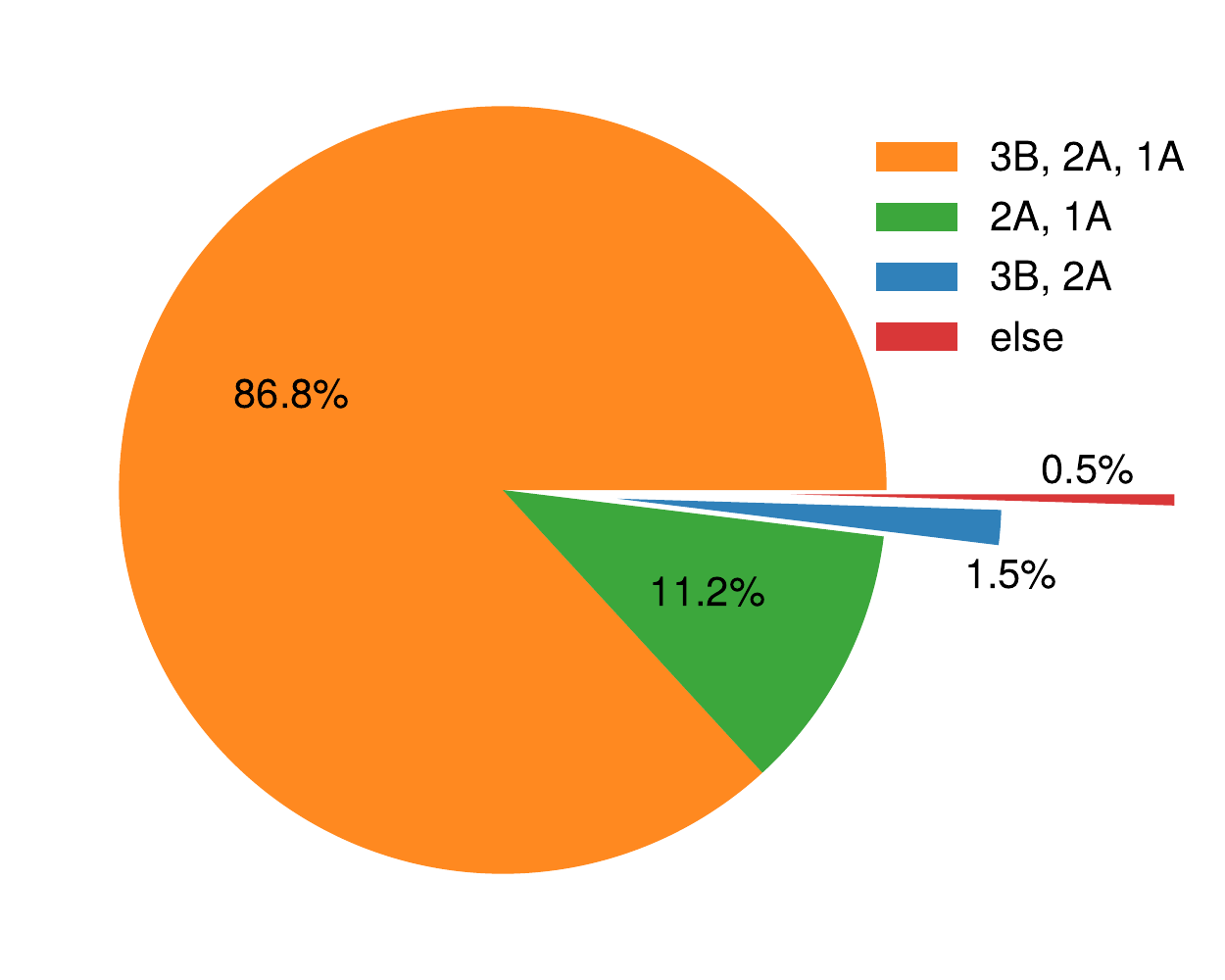}
\caption{Fraction of each combination of bond states in dissociation processes.}
\label{fig:bond}
\end{figure}

In the main text, we classify molecular bonds into 12 states (Table~1 in the main text) and report that more than \SI{99}{\percent} of the bonds dissociate in a specific pathway of states.
Here, we discuss this analysis in detail.
In the analysis, we checked the states of each bond in logarithmically spaced time intervals.
We determined that the molecular bond had passed through the state if the state appeared at least once during the time evolution.
The order and number of appearances were not taken into account.
We carried out this analysis for all molecular pairs in trajectories of $T = 0.491$ and obtained the fraction of each set of states (Fig.~\ref{fig:bond}).
More than \SI{80}{\percent} of the molecular pairs visited the states ``3B'', ``2A'', and ``1A'' in the trajectories; \SI{12.8}{\percent} of the pairs did not visited the states ``3B'' or ``1A'' but visited the others.
As mentioned before, the analysis was conducted in logarithmically spaced times, therefore we consider all three patterns of appearances as a single path of states shown in Fig.~4 in the main text.
We also observed that \SI{0.07}{\percent} of all bonds did not dissociate in any of the three states ``3B'', ``2A'', and ``1A''.

\subsection*{Inherent dynamics}
This section provides information on numerical procedures to obtain the data of the Kob-Andersen (KA) system in Fig.~5 in the main text.
The molecular dynamics (MD) simulations in the \textit{NVT} ensemble using the Nos\'{e}-Hoover thermostat were performed for equilibration runs, and subsequently, the MD simulations in the \textit{NVE} ensemble were performed for production runs.
We consider $N = 1000$ particles enclosed in a square box with periodic boundary conditions.
The number density is $\rho = N/L^3 = 1.204$.
We use identical interaction parameters to those of the original paper by Kob and Andersen~\cite{Kob_Andersen_PRL_1994}.
We ensure the continuity at the cutoff length $r_\text{c} = 2.5\sigma_{ij}$ up to the first derivative of the potential as described in the main text.
The number of samples is 3800.
The relaxation time is measured with $F_s(k = 7.25, \tau_\alpha) = e^{-1}$.

\subsection*{Time-dependent variance of potential energies}
In the main text, we introduce the time-dependent variance of potential energies $V(t) = \Var\bqty{E_\text{IS}(t) - E_\text{IS}(0)}$ to quantify the size of the phase space explored by IS trajectories.
$E_\text{IS}(t)$ is the potential energy of a configuration in an IS trajectory at time $t$.

\begin{figure}
\centering
\includegraphics[width=\linewidth]{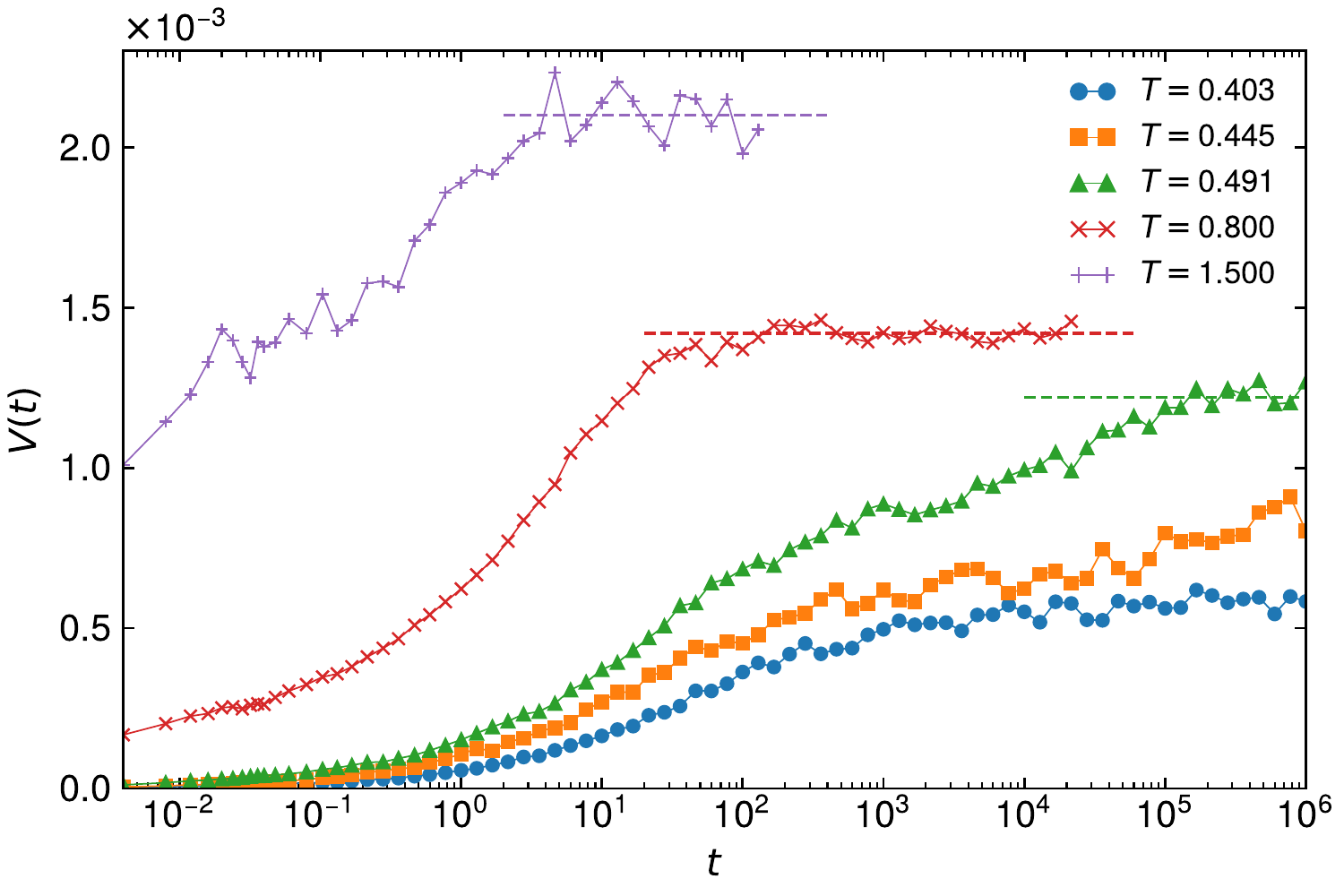}
\caption{$V(t)$ for various temperatures.
Horizontal lines with each color represent the long-time limit $V_\infty$.}
\label{fig:Vt}
\end{figure}

Figure~\ref{fig:Vt} shows $V(t)$ for various temperatures.
Horizontal lines are the long-time limits $V_\infty$, which was estimated from $V(t)$ at $t \gtrsim \tau_\alpha$.
For temperatures $T = 0.403, 0.445, 0.491$, we used configurations generated by the parallel tempering method as initial configurations.
The number of initial configurations is 95, and the number of the isoconfigurational trajectories is 6 for $T = 0.403$ and $T = 0.445$ and 24 for $T = 0.491$.
For $T = 1.500, 0.800$, we generated initial configurations by the MD simulation in the \textit{NVT} ensemble using the Nos\'{e}-Hoover thermostat and subsequent \textit{NVE} runs were performed.
The number of trajectories is 1000 for $T = 1.500$ and 5600 for $T = 0.800$.
For these temperatures, the simulation time is sufficiently long that $V(t)$ converges to the long-time limit $V_\infty$.
At a short time scale, $V(t)$ at $T = 1.500$ and $0.800$ does not start from 0.
The reason is that even one MD step shifts the system to a different basin of attraction at these high temperatures.
On the other hand, at low temperatures, $V(t)$ continuously deviates from 0 because the system does not leave the initial basin in just one MD step.
A similar observation was made from the overlap function of IS trajectories of the KA system~\cite{Baity-Jesi_2021}.

\subsection*{Correlation between low-frequency vibrational modes and relaxations in the Kob-Andersen system}
\begin{figure}
\centering
\includegraphics[width=\linewidth]{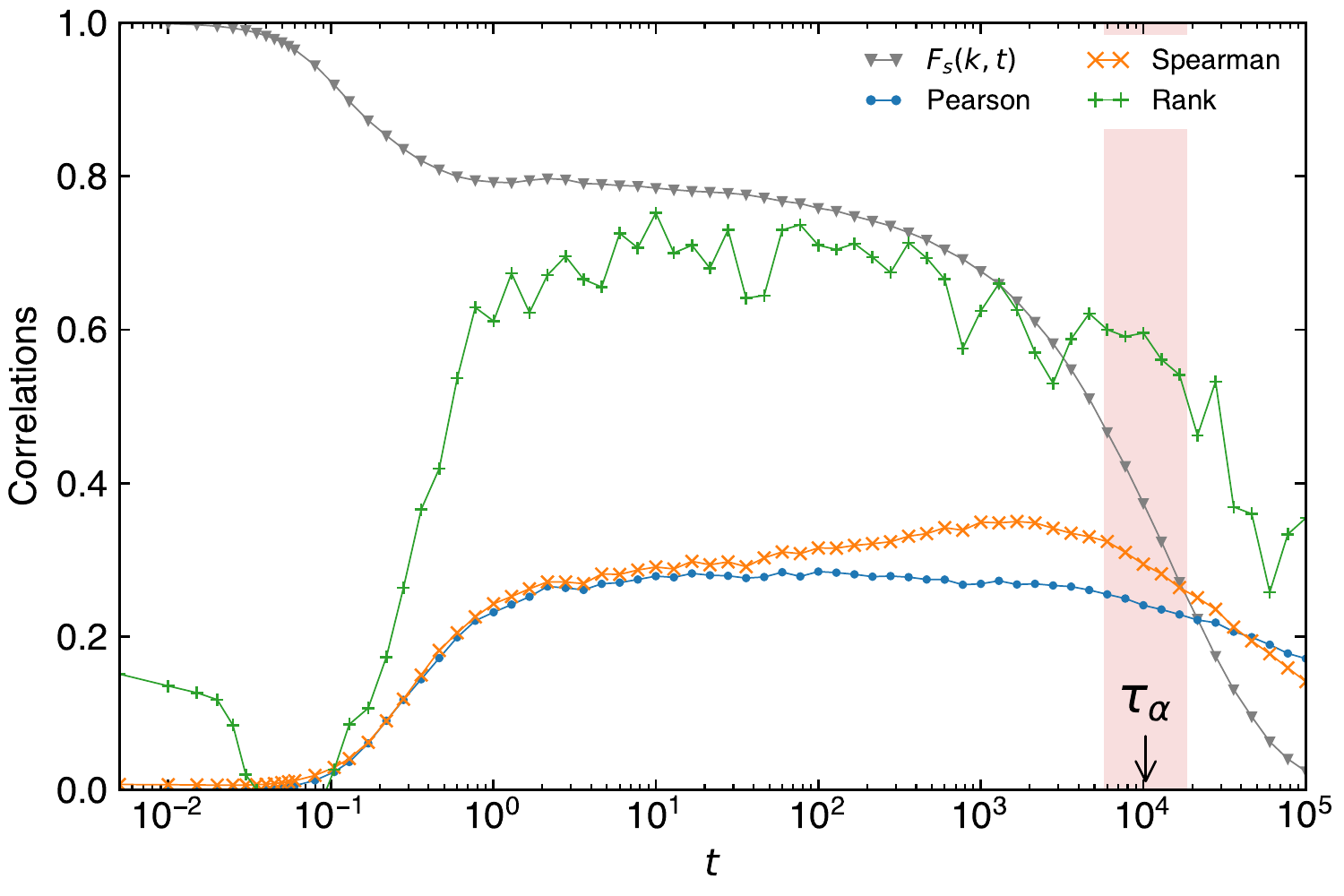}
\caption{The Rank, Pearson, and Spearman correlations between the low-frequency modes and the propensity of motion in the KA system ($T = 0.4308$).}
\label{fig:predictKA}
\end{figure}

We carried out a parallel tempering simulation to equilibrate the KA system in $T = 0.4308$.
The parameters and setups are identical to the above.
In our MD simulations, 18 replicas were used, and each replica corresponded to a temperature from 0.7800 to 0.3971.
The MD simulations were performed in the \textit{NVT} ensemble using the Nos\'{e}-Hoover thermostat with a time step of 0.005.
Exchange trials were performed every 2800 MD steps using the Metropolis criterion.
Sampling was started after 80000 exchange trials and was performed every 10000 trials thereafter.
We checked the equilibration by the absence of aging in $F_s(k,t)$.
Using these configurations as initial positions, we carried out the simulations in the \textit{NVE} ensemble for production runs.
The number of initial configurations is 72, and the number of the isoconfigurational ensemble is 5.
We used the norms of eigenvectors of each particle summed over the lower \SI{1}{\percent} of all modes (i.e.\ the 30 lowest-frequency modes) at IS corresponding to the initial configuration as the predictor and the propensity of motion $\abs{\vb*{r}_i(t) - \vb*{r}_i(0)}$ as the actual relaxations.
For the definitions of the evaluators (the Rank, Pearson, and Spearman correlations), see the main text.
The correlation persists to the timescale of $\alpha$ relaxation in Fig.~\ref{fig:predictKA}.

\bibliography{my_bibfile.bib}
\end{document}